\documentclass[pra,aps,twocolumn,10pt,showpacs,nofootinbib]{revtex4-2}
\usepackage[utf8]{inputenc}
\usepackage[T1]{fontenc}
\usepackage{amsmath,amsthm,charter,amssymb,amsfonts}
\usepackage{mathcmd}
\usepackage{graphicx} 

\usepackage{braket}
\usepackage{bm,bbold}
\usepackage{mathrsfs}
\usepackage{stmaryrd}
\usepackage{esint}
\usepackage{xcolor}

\usepackage{tikz}
\usepackage{rotating} 
\usepackage{float}

\usepackage{ulem}
\usepackage[colorlinks=true, linkcolor=purple, citecolor=blue]{hyperref}

\newcommand{\pb}{\mathrm{p}}
\newcommand{\Tr}{\mathrm{Tr}}

\begin{document}

\title{Retrodiction of measurement outcomes on a single quantum system reveals entanglement with its environment}

\author{Julien Pinske}
     \email{julien.pinske@nbi.ku.dk}
     \affiliation{Niels Bohr Institute, University of Copenhagen, Blegdamsvej 17, DK-2100 Copenhagen, Denmark}

\author{Klaus M\o lmer}
    \affiliation{Niels Bohr Institute, University of Copenhagen, Blegdamsvej 17, DK-2100 Copenhagen, Denmark}

\date{\today}

 \begin{abstract}
The density matrix yields probabilistic information about the outcome of measurements on a quantum system, but it does not distinguish between classical randomness in the preparation of the system and entanglement with its environment. 
Here, we show that retrodiction, employing both prior and posterior knowledge, gives rise to conditional probabilities for measurements on a single system, that can witness if it is part of a larger composite system. The degree of certainty with which one can retrodict the outcomes of multiple measurements on a system can witness both the existence and the quantitative nature of its entanglement with the environment. 
\end{abstract}
    
    \maketitle

    \begin{figure*}[t]
        \centering
        \begin{tikzpicture}
        \node at (-8.9,0.13) {\includegraphics[width=0.42\textwidth]{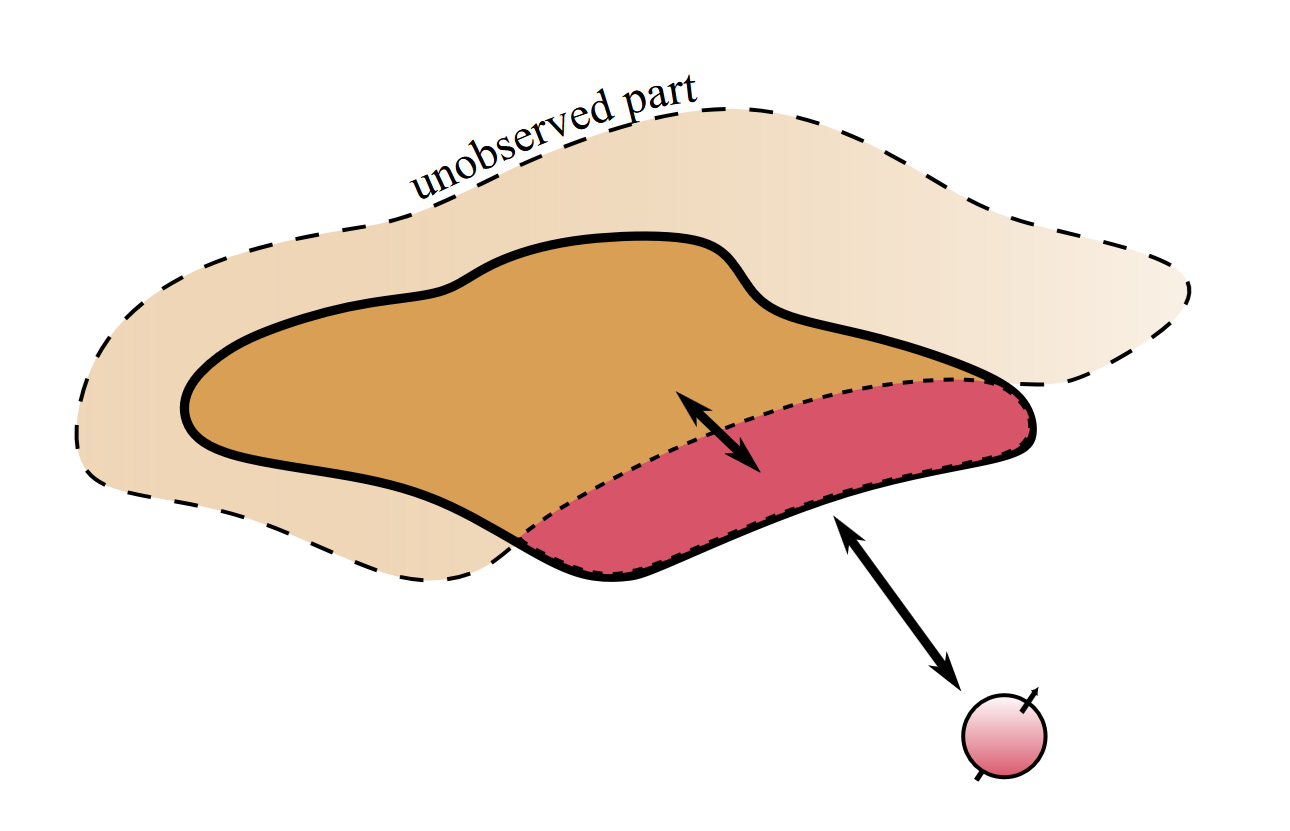}};
        \node at (-7.89, -1.2) {$\Omega_{m}$};
        \node at (-6.23, -1.85) {$m$};
        \node at (-9.16, -1.1) {$\rho_A$};
        \node at (-9.56, 0.4) {$B$};
        \node at (-7.96, -0.05) {$A$};
        \node at (-12,2.2) {(a)};
        \node at (-0.2,0.2) {\includegraphics[width=0.5\textwidth]{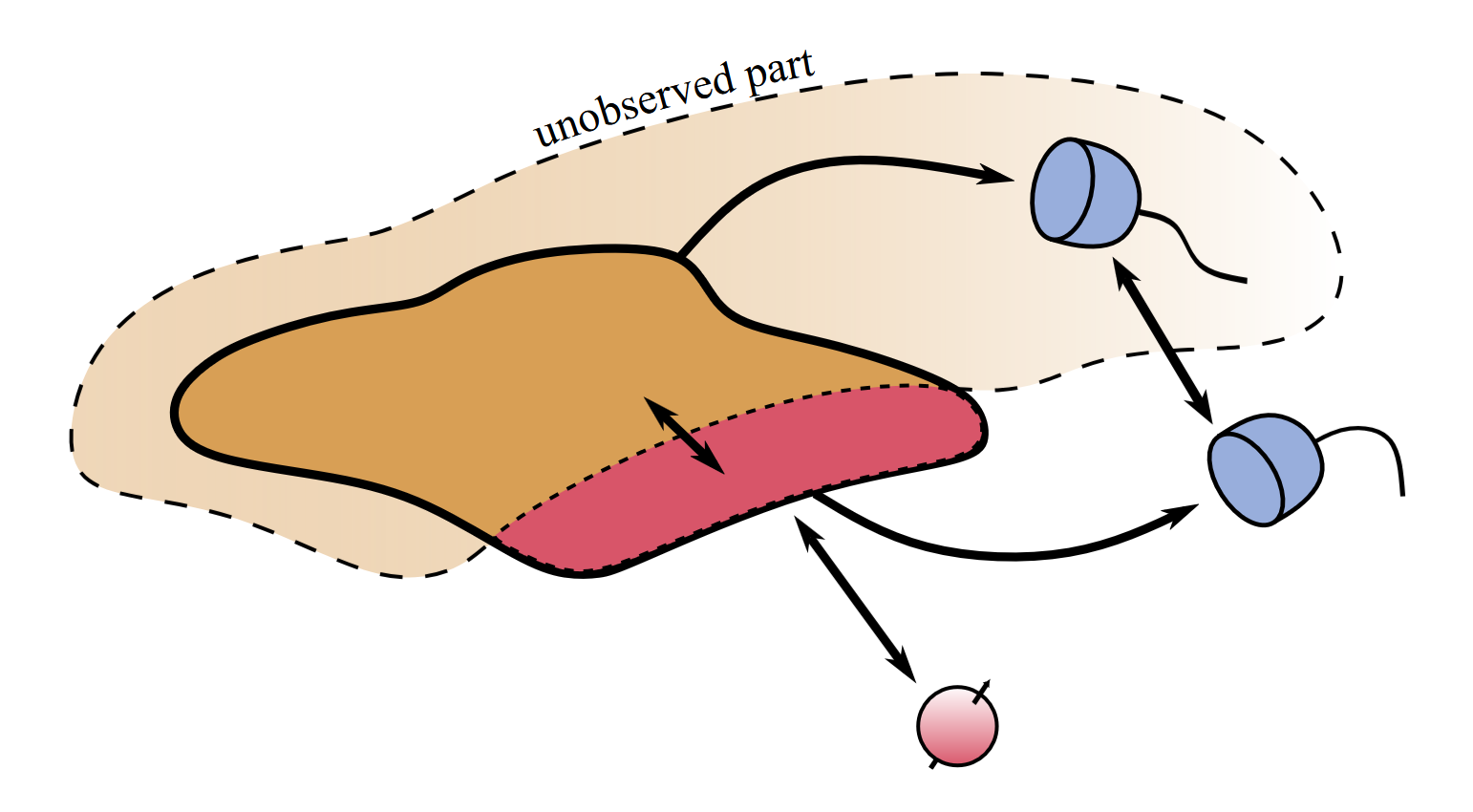}};
        \node at (-1.2, -1.1) {$\rho$};
        \node at (-1.6, 0.4) {$B$};
        \node at (0, -0.05) {$A$};
        \node at (3, 0.4) {$E$};
        \node at (0.07, -1.2) {$\Omega_{m}$};
        \node at (1.73, -1.85) {$m$};
        \node at (-4,2.2) {(b)};
        \end{tikzpicture}
        \caption{\label{fig:setup} Verifying the existence of entanglement within a composite quantum system $AB$.
        Measurements performed on subsystem $A$ with outcomes $m$ are associated with the application of the operator $\Omega_{m}$.  
        (a) The composite system $AB$ is prepared in the initial state $\rho$, and the probabilities for the outcomes $m$ are determined by the reduced density matrix $\rho_A=\Tr_B(\rho)$, allowing no conclusions about the entanglement with $B$.
        (b) If the composite system is initially prepared in $\rho$ and is ultimately detected  as described by a measurement effect $E$, the 
        retrodicted, or postselected, probabilities for the outcomes $m$ on $A$ can reveal entanglement, imposed by $\rho$ and $E$  within the composite system.}
    \end{figure*}
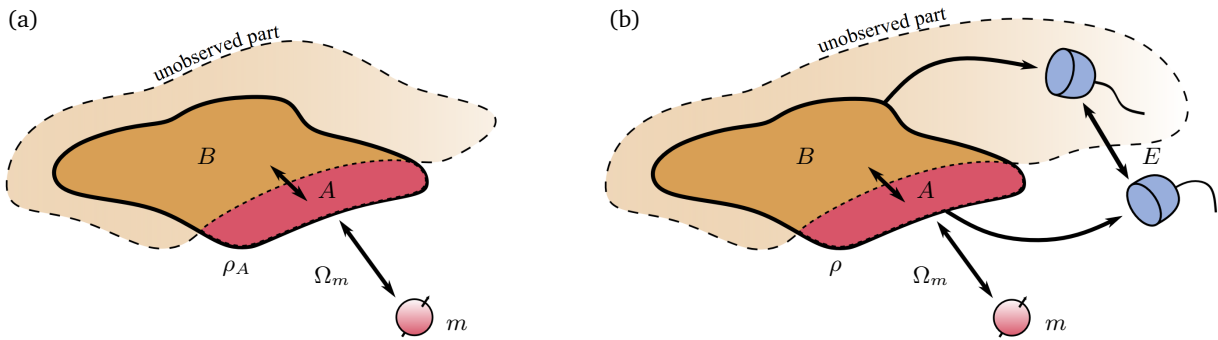

    \section{Introduction}
    \label{sec:intro}

In the treatment of open quantum systems, a system $A$ experiences loss of excitation and coherence, due to its interaction and entanglement with its environment $B$. 
The probabilities of the outcomes of measurements on the whole system $AB$ are governed by the density matrix $ \rho_{AB}$, while for measurements on the smaller system, they depend only on the reduced density matrix $\rho_A=\Tr_B(\rho_{AB})$. Similarly, entangled states shared by users of a quantum network appear mixed, when local measurement outcomes are not communicated between the parties involved, and approaches to determine global properties need access to joint correlations \cite{T00,VP09,H24}.

The situation is different if we have access to both the initial density matrix $\rho_{AB}$ and the result of a posterior joint measurement on the composite system $AB$. 
This situation is associated with the process of postselection, and for the case of pure initial and final states, separate expressions exist  for the outcome of weak \cite{AAV88} and projective measurements \cite{ABL64}. 
In a more general setting, the probability of the outcome of any kind of measurement at any time may be conditioned on the initial density matrix, the known Hamiltonian and dissipative evolution of the system and the results of any sequence of earlier and later measurements on the system \cite{GJM13,ZM17}. 
This includes, for example, the monitoring of an open quantum system, such as continuous homodyning by photodetection \cite{W02}, and the estimation of unknown classical perturbations, applied to a probed quantum system \cite{T09,T11}.

Even though we only address measurements on $A$, the retrodicted outcome probabilities are not captured by the reduced density matrix $\rho_A$ but by the full density matrix $\rho_{AB}$ and the results of later observations on the system $AB$. 
An illustration of this phenomenon is the remarkable fact that it is possible to pass a single spin-1/2 particle to an experimentalist, let them do an experiment and return the particle for subsequent inspection, and tell with certainty what was the outcome of the measurement of any of the three Cartesian spin components $\sigma^x$, $\sigma^y$, and $\sigma^z$ \cite{VAA87}. 
For two spin components such retrodiction is simple: first provide an eigenstate of $\sigma^x$ and measure $\sigma^y$ of the returned particle, then the intermediate measurement of $\sigma^x$ or $\sigma^y$ would have to agree with the prior or posterior value, respectively. 
The retrodiction of three candidate measurement outcomes is possible when the spin is initially entangled with another spin, and they are finally measured to be in an entangled state \cite{VAA87,SSW03}.

Although the full density matrix $\rho_{AB}$, and a complete account for the postselection by joint measurements on $A$ and $B$ seem necessary for retrodiction of measurements on system $A$, we show that it is possible to capture the same information using a statistical operator acting only on the small system. 
This offers a picture where the entanglement between the system and its environment is witnessed by inseparable correlations between state matrices and measurement effects on the probe system.
Our analysis suggests that free (distillable) entanglement \cite{BBP96} between $A$ and $B$ is necessary for the sharp retrodiction of multiple noncommuting observables.

Beyond verifying entanglement, we are also interested in determining the dimension of the external (inaccessible) system.
To address this issue in a systematic way, we appropriate the concept of a dimension witness for postselected states.
Originally applied to Bell inequalities \cite{BPA08,JP11,PM20}, dimension witnesses were employed in prepare-and-measure scenarios \cite{BNV13,CCB17} and in the analysis of experimental data \cite{HGM12,ABC12,TSJ24}.
Existing approaches to witnessing the dimension of the environment of a system require joint measurements to go beyond the predictions offered by the reduced density matrix. 
The theory developed here for postselected or retrodicted outcomes bypasses this limitation. 
We note that questions concerning the discernibility of different ensemble realizations of a reduced density matrix was recently discussed from a similar retrodiction perspective  \cite{LBS25}. 

The article is structured as follows.
Section~\ref{sec:past-state} provides a brief review of the past quantum state formulation of retrodiction and postselection. 
As our first key result, we derive a reduced operator for the accessible part of the system in Sec.~\ref{sec:red-past-state}.
It is shown that this operator's positivity yields a separability criteria.
In Sec. \ref{sec:CV-retro}, we apply the reduced operator to continuous-variable systems and show how its phase-space representation can be used to efficiently calculate outcome probabilities.
Section~\ref{sec:dim-wit} introduces dimension witnesses for past states, the second key result of this paper.
In Sec.~\ref{sec:examples} the theory is applied to measurements of mutually unbiased bases.
Finally, Sec.~\ref{sec:Fin} provides a summary of the article and concluding remarks.

\section{Past quantum states}
\label{sec:past-state}

Here, we review the notion of past quantum states~\cite{GJM13}.
Let $\rho$ be a density matrix accounting for prior knowledge about a system.
A measurement (POVM \cite{W18}) is described by a collection of operators $\Omega_m$ satisfying $\Omega_m^\dag \Omega_m\geq 0$ and $\sum_m\Omega_m^\dag \Omega_m=\mathbb{1}$, where $\mathbb{1}$ is the identity and $m$ labels different outcomes.
The probability for obtaining $m$ is given by Born's rule $p(m)=\Tr(\Omega_m\rho\Omega^\dag_m)$ and the state after the measurement reads $\Omega_m\rho\Omega_m^\dag/p(m)$.
     
If a second measurement, described by operators  $\Lambda_n$, is performed, we iterate the above formalism and find that the joint probability of obtaining first outcome $m$ and then outcome $n$, is $p(m,n) = 
\Tr(\Lambda_n\Omega_m\rho \Omega^\dag_m\Lambda^\dag_n)$.
With this joint probability at hand, we can infer the probability that the early measurement yielded outcome $m$ conditioned on any definite outcome $n^\prime$: $p(m|n^\prime)=p(m,n^\prime)/\sum_{m^\prime} p(m^\prime,n^\prime)$.  
The knowledge of the late measurement result $n^\prime$ causes a modification of Born's rule,
    \begin{equation}
        \label{eq:Bayes}
        p_{\pb}(m)=\frac{\Tr(\Omega_m\rho\Omega_m^\dag E)}{\sum_{m^\prime}\Tr(\Omega_{m^\prime}\rho\Omega_{m^\prime}^\dag E)},
    \end{equation}
where we denote $E=\Lambda_{n^\prime}^\dagger \Lambda_{n^\prime}$ as the effect matrix. 
Without posterior information, i.e., $E=\mathbb{1}$, Eq.~\eqref{eq:Bayes} reduces to Born's rule.

This equation applies more generally, incorporating all measurements as well as the unitary and dissipative dynamics occurring before and after the $\Omega_m$ measurement until a final time $T$. 
Then, the matrices $\rho(t)$ and $E(t)$, both available at $T$, provide the probability of the outcome of any measurement on the system at an earlier time $t\leq T$.
Following Ref. \cite{GJM13}, we thus refer to the pair of matrices $(\rho,E)$ as the past quantum state \cite{GJM13}. 

For the projective measurement of an observable $M$, Eq. \eqref{eq:Bayes} applies with $\Omega_m=\Pi_m$ denoting orthogonal projection operators on the eigenstates of $M$.
Past mean values $\braket{M}_\pb$ can then be calculated in a familiar manner from the conditional probabilities in Eq. \eqref{eq:Bayes}, i.e., $\braket{M}_\pb=\sum_m \lambda_m p_{\pb}(m)$, with $\lambda_m$ denoting the eigenvalues of $M$.
Similarly, variances $\mathrm{Var}_\pb(M)=\braket{M^2}_\pb-\braket{M}_\pb^2$ can be determined. 
As illustrated in Sec.~\ref{sec:intro}, outcome probabilities and variances assigned by retrodiction do not obey Heisenberg’s uncertainty relation \cite{H04,BJD20}. 
   
\subsection{Past quantum state of a composite system}

Throughout this article, we are interested in the following situation. Consider a two-party quantum system $AB$ with states in the Hilbert space $\mathcal{H}_A\otimes \mathcal{H}_B$. 
An experimentalist, having only access to system $A$, performs a number of local measurements described by sets of $\Omega_{m}$ acting on $\mathcal{H}_A$. 
Conditioned by the outcome of measurements on the joint system, illustrated by the two detectors in Fig. \ref{fig:setup}, the local measurement results are retained (postselected) and the conditional probability $p_{\pb}(m)$ is estimated according to the frequency of occurrence of the outcome $m$.
Given these outcomes of measurements of possibly multiple different properties of $A$, is it possible to verify that system $A$ under investigation is part of a composite system $AB$? 
    
To answer this question, we first note that when $\rho$ and $E$ refer to a larger composite system $AB$, the measurement outcomes on $A$ have the probabilities, cf., Eq. \eqref{eq:Bayes},
    \begin{equation}
        \label{eq:Bayes-local}
        p_{\pb}(m)=\frac{\Tr\{(\Omega_{m}\otimes\mathbb{1})\rho(\Omega_{m}^\dag\otimes\mathbb{1}) E\}}{\sum_{m^\prime}\Tr\{(\Omega_{m^\prime}\otimes\mathbb{1})\rho(\Omega_{m^\prime}^\dag\otimes\mathbb{1}) E\}}.
    \end{equation}
If the measurement effect is trivial, $E=\mathbb{1}^{\otimes2}$, outcomes are determined by the reduced density matrix $\rho_A$ and any entanglement with $B$ merely produces a mixed state. 
Then, it is not possible to distinguish the presence of system $B$ from classical randomness causing a similar mixed-state character.
 
\section{Reduced past states}
\label{sec:red-past-state}

In general, given a past state $(\rho,E)$ on the composite system $AB$, it is not possible to assign individual reduced operators $\rho_A$ and $E_A$ that reproduce the probabilities in Eq. \eqref{eq:Bayes-local} \cite{SGB14}.
Nevertheless, we will show that it is possible to introduce a statistical (state-like) operator on subsystem $A$, that combines information from both $\rho$ and $E$, and thus fully represent Eq.~\eqref{eq:Bayes-local}.

First, we expand both $\rho$ and $E$ in terms of positive (semi-definite) product operators acting on $\mathcal{H}_A\otimes\mathcal{H}_B$,
    \begin{equation}
        \label{eq:prod-exp}
	    \rho=\sum_j t_j \rho^j\otimes \omega^j,\quad E=\sum_k s_k E^k\otimes F^k.
    \end{equation}
This expansion is always possible by taking a basis of Hermitian tensor-product operators acting on $\mathcal{H}_A\otimes \mathcal{H}_B$ and then spectrally decomposing each operator into a linear combination of (positive) eigenspace projectors \cite{W18}.
However, we note that if $\rho$ ($E$) is entangled, then at least one of the coefficients $t_j$ ($s_k$) must be negative.
Inserting the above expansion into Eq. \eqref{eq:Bayes-local} leads to
    \begin{equation}
	       p_{\pb}(m)\propto\sum_{j,k} x_{jk} \Tr(\Omega_m\rho^j\Omega_m^\dag E^k),
    \end{equation}
where $x_{jk}=t_j s_k \Tr(\omega^j F^{k})$. This equation suggests that in the presence of entanglement with other degrees of freedom, a quantum  system can be described by a reduced past state, which takes the form of a collection of past quantum states $(\rho^j, E^k)$, with weight factors $x_{jk}$. 

The reduced past quantum state can be conveniently represented by the Hermitian operator
    \begin{equation}
	\label{eq:red-past-state}
	   \begin{split}
	       \Xi_A &=\sum_{j,k}x_{jk}\rho^j \otimes E^k\\
	   \end{split}
    \end{equation} 
    acting on $\mathcal{H}_A\otimes\mathcal{H}_A$.
    Despite its appearance, which suggests that $\Xi_A$ must be inferred from the specific expansion in Eq. \eqref{eq:red-past-state}, it is possible to write $\Xi_A$ in a form that does not require such an expansion, 
\begin{equation}
    \label{eq:red-past-state-general}
	   \begin{split}
	       \Xi_A &=\Tr_B\{(\rho_{AB}\otimes \mathbb{1}_{A})(\mathbb{1}_{A}\otimes E_{BA})\}.
	   \end{split}
\end{equation}
The statistical operator $\Xi_A$, acting on $\mathcal{H}_A\otimes\mathcal{H}_A$, describes our knowledge of measurement outcomes on system $A$ while capturing the role of the full $\rho$ and $E$, including entanglement within the larger system $AB$. 
In particular, the probabilities in Eq. \eqref{eq:Bayes-local} can be obtained from $\Xi_A$,
    \begin{equation}
        \label{eq:prob-Xi}
        p_{\pb}(m)=\frac{\Tr(\mathcal{M}_m(\Xi_A))}{\sum_{m^\prime}\Tr(\mathcal{M}_{m^\prime}(\Xi_A))},
    \end{equation}
where $\mathcal{M}_m(\Xi_A)$ is a linear map describing the measurement $\Omega_m$, defined by its action on product operators $\mathcal{M}_m(\rho^j\otimes E^k)=\Omega_m \rho^j \Omega_m^\dag E^k$.
For $E=\mathbb{1}^{\otimes 2}$ we recover the reduced density matrix $\Xi_A=\rho_A\otimes\mathbb{1}$, with $\rho_A=\Tr_B(\rho)=\sum_j t_j \rho^j$. 

\subsection{Separability of the reduced past state}
\label{ssec:sep-red-past-state}

In general, if both $\rho$ and $E$ are separable across $AB$, then there exists an expansion as in Eq. \eqref{eq:prod-exp}, such that $t_j,s_k\geq 0$, and thus $x_{jk}\geq 0$.
From the perspective of system $A$,  separability of $\rho_{AB}$ and $E_{AB}$ implies separability of the matrix $\Xi_A$. 
It is tempting to think of this as a temporal separability---between the preparation and the measurement effect.
Regarded as a state matrix, $\Xi_A$ is in that case a convex combination of (positive) product operators and is itself a positive operator on $\mathcal{H}_A\otimes\mathcal{H}_A$.
In contrast, for a general past state $(\rho,E)$, some of the coefficients $x_{jk}$ can become negative and $\Xi_A$ does not need to be positive.
In particular, if $\Xi_A$ has a negative eigenvalue, either $\rho$ or $E$ must be entangled.

As an example that illustrates this behavior, consider both $\rho$ and $E$ to be given by a projection onto the maximally entangled state $\ket{\phi^+}=\sum_{j}\ket{jj}$.
Using Eq. \eqref{eq:red-past-state-general} we find that $\Xi_A$ corresponds to the swap operator, $\Xi_A=\sum_{j,k}\ket{jk}\bra{kj}$, which is not positive.

On the other hand, positivity of the statistical operator $\Xi_A$ is a necessary, but not sufficient condition, for separability of the past state $(\rho,E)$.
To see this, consider an ensemble representation of the effect matrix $E$, i.e., $E=\sum_x q_x \ket{e_x}\bra{e_x}$, with probabilities $q_x\geq 0$.
Further note that any (pure) effect $\ket{e_x}$ can be written as $\ket{e_x}=(M_x\otimes \mathbb{1}_B)\ket{\phi^+}$, for some (local) operator $M_x$ acting on $\mathcal{H}_A$.
Then, Eq. \eqref{eq:red-past-state-general} takes the form
\begin{equation}
    \label{eq:bound-past-state}
    \Xi_A=\sum_x q_x (\mathbb{1}\otimes M_x)\rho^\Gamma (\mathbb{1}\otimes M_x^\dag),
\end{equation}
where $\Tr_B\{(\rho_{AB}\otimes\mathbb{1}_A)(\mathbb{1}_A\otimes \ket{\phi^+}\bra{\phi^+}_{BA})\}=\rho^\Gamma$ denotes the partial transpose \cite{P96} of $\rho$.
Note that if $M_x$ is unitary, $M_x^\dag M_x=\mathbb{1}$, then $\Tr(\Xi_A)=\Tr(\rho^\Gamma)=1$, because transposition preserves the trace of a matrix and $\sum_x q_x=1$.

If the density matrix $\rho$ has a positive partial transpose (PPT) $\rho^\Gamma\geq 0$, then the statistical operator $\Xi_A$ is positive as well.
We note that for low-dimensional systems, e.g., $\mathcal{H}_A\otimes\mathcal{H}_B=\mathbb{C}^2\otimes\mathbb{C}^3$, the matrix $\rho$ is PPT if and only if $\rho$ is separable \cite{HHH96}.
More interestingly, this also shows that PPT states (and effects) lead to $\Xi_A$ being positive.

All PPT states are either separable or bound entangled, that is they cannot be distilled into pure-state entanglement using local operations and classical communication \cite{HHH98,PMM04}.
It remains an open problem whether there exist bound entangled states which are not PPT \cite{DSS00,HRZ22}.
Since, negative eigenvalues of $\Xi_A$ are necessary for explaining certain postselected outcomes, e.g., the joint certain retrodiction of three Cartesian spin components (cf. Sec. \ref{sec:intro}), we conjecture that bound entangled states are not useful for certain retrodiction of multiple noncommuting observables.

\subsection{Criteria for prior and posterior entanglement}

Entanglement in either $\rho$ or $E$ can manifest itself in a negative eigenvalue of the statistical operator $\Xi_A$.
The question arises whether we can assign the origin of this negativity to either $\rho$ or $E$ being entangled across $AB$.
Decomposing $\rho$ and $E$ as in Eq. \eqref{eq:prod-exp} leads to $\Xi_A$ taking the form in Eq. \eqref{eq:red-past-state}.
We obtain reduced operators by tracing out the first and second $A$ system, respectively, 
\begin{equation}
    \begin{split}
        \Tr_{2}(\Xi_A) & = \sum_j t_j p_j \rho^j,\quad 
        \Tr_{1}(\Xi_A)  = \sum_k s_k q_k E^k.\\
    \end{split}
\end{equation}
Here we defined $p_j=\Tr(\omega^j E_B)$ and $q_k=\Tr(\rho_B F^k)$ depending
on the reduced operators 
\begin{equation}
    \begin{split}
        \rho_B & = \Tr_A(\rho) = \sum_j t_j \omega^j,\\
        E_B & = \Tr_A(E) = \sum_k s_k F^k,\\
    \end{split}
\end{equation}
respectively.
Since $\omega^j,E_B\geq 0$ are positive operators, we have that $p_j\geq 0$.
Then, any negative eigenvalue in the operator $\Tr_{2}(\Xi_A)$ must originate from a decomposition \eqref{eq:prod-exp} where one of the coefficients $t_j$ is negative.
This is then a sufficient criteria for $\rho$ to be entangled across $AB$.
Likewise, from $q_k\geq 0$, it follows that a negative eigenvalue of $\Tr_{1}(\Xi_A)$ implies that $E$ is entangled.

To see that these are not necessary criteria for entanglement in the individual operators $\rho$ and $E$, consider again $\rho$ and $E$ to be in the maximally entangled state $\phi^+$ (Sec. \ref{ssec:sep-red-past-state}).  
In this case, $\Xi_A=\sum_{j,k}\ket{jk}\bra{kj}$ is the swap operator.
The reduced operators $\Tr_1(\Xi_A)=\Tr_2(\Xi_A)=\mathbb{1}_A$ correspond to the unit matrix, which is obviously positive.
Entanglement is present in $\rho$ and $E$, and the statistical operator $\Xi_A$ verifies this entanglement via its negative eigenvalue.
However, the reduced operators are positive, and thus it is not revealed whether the negativity originates from $\rho$ or $E$ (or both) being entangled.

\subsection{Retrodiction of spin components}

    As a first example, we consider the joint system $AB$ to consist of two qubits and let
    \begin{equation}
        \label{eq:VAA-state}
        \begin{split}
            \ket{\psi}&=\frac{1}{\sqrt{2}}(\ket{00}+\ket{11}),\\
            \ket{e}&=\frac{1}{\sqrt{2}}\ket{00}+\frac{1}{2}\big(e^{i\pi/4}\ket{10}+e^{-i\pi/4}\ket{01}\big),\\
        \end{split}
    \end{equation}
    constitute the initial and postselected state. 
    As shown in \cite{VAA87}, in this case it is possible to assign with certainty the outcome of any of the three Pauli measurements, as $\braket{\sigma^x}_{\pb}=\braket{\sigma^y}_{\pb}=\braket{\sigma^z}_{\pb}=1$, on the first qubit $A$.
    The reduced past state is readily obtained from Eq. \eqref{eq:red-past-state-general},
    \begin{equation}
        \begin{split}
            \Xi_A&=\ket{e}\bra{e}^\Gamma,\\
        &=\frac{1}{4}(\ket{10}\bra{10} + \ket{01}\bra{01}) + \ket{\zeta}\bra{00} + \ket{00}\bra{\zeta},
        \end{split}
    \end{equation}
    where we defined
    \begin{equation}
        \ket{\zeta}=\frac{1}{4}\ket{00}+\frac{1+i}{4}(\ket{10} + \ket{01}) + \frac{i}{4}\ket{11}.
    \end{equation}
    The reduced past state $\Xi_A$ has a negative eigenvalue $\lambda_{\min}=-1/4$, the signature of non-separability of $\Xi_A$ and evidence of entanglement between $A$ and $B$.

\section{Retrodiction in continuous-variable systems}
\label{sec:CV-retro}

    In this section, we employ the reduced past state $\Xi_A$ for the retrodiction of measurement outcomes in continuous-variable quantum systems \cite{WPG12}.
    These include quantized bosonic modes related to the different degrees of freedom of the electromagnetic field. 

    \subsection{Husimi representation of past quantum states}

    In quantum optics it has been of interest to measure the complex field amplitude $\alpha$ using heterodyne detection.
    This is described by mapping onto coherent states $\Omega_{\alpha}=\tfrac{1}{\sqrt{\pi}}\ket{\alpha}\bra{\alpha}$, which form an overcomplete basis, $\tfrac{1}{\pi}\int \ket{\alpha}\bra{\alpha}d^2 \alpha=\mathbb{1}$ \cite{VW06}, and $d^2\alpha=d\mathrm{Re}(\alpha)d\mathrm{Im}(\alpha)$ denotes integration over the real and imaginary part of $\alpha$.
    In this case, system $A$ corresponds to a mode of the electromagnetic field and we can identify the outcome probability in Eq.~\eqref{eq:Bayes} with the product of the two Husimi-functions $Q_{\rho}(\alpha)=\tfrac{1}{\pi}\braket{\alpha|\rho|\alpha}$ and $Q_{E}(\alpha)=\tfrac{1}{\pi}\braket{\alpha|E|\alpha}$,
    \begin{equation}
        \label{eq:prob-Husimi}
        p_{\pb}(\alpha)=\frac{Q_{\rho}(\alpha)Q_{E}(\alpha)}{\int  Q_{\rho}(\alpha)Q_{E}(\alpha) d^2\alpha}.
    \end{equation}
    
    As we pointed out in Sec. \ref{sec:red-past-state}, the form of the probabilities $p_{\pb}(\alpha)$ in Eq.~\eqref{eq:prob-Husimi} assumes that the state of mode $A$ is not entangled with another mode $B$.
    For the general situation, a description in terms of the operator $\Xi_A$ is necessary to account for the initial entanglement in the joint system $AB$, as well as the entangling effect of the postselection. 
    The reduced past state $\Xi_A=\sum_{j,k}x_{jk}\rho^j\otimes E^k$, expanded as in Eq. \eqref{eq:red-past-state}, then leads to
    \begin{equation}
        \label{eq:gen-Husimi}
        p_{\pb}(\alpha)\propto \sum_{j,k}x_{jk}Q_{\rho^j}(\alpha)Q_{E^k}(\alpha),
    \end{equation}
    with corresponding Husimi functions of both the states and effects in the ensemble $(\rho^j,E^k)$.

    \subsection{Wigner representation of past quantum states}

    While the Husimi representation is naturally associated with measurements of the field amplitude, the Wigner representation
    \begin{equation}
        \label{eq:def-Wigner}
        W_\rho(\alpha)=\frac{2}{\pi}\Tr(D(\alpha)(-1)^{\hat{a}^\dag \hat{a}}D(\alpha)^\dag \rho),
    \end{equation}
    of a density matrix $\rho$, is intimately tied to the homodyne measurement of the phase-dependent quadratures $\hat{x}_\varphi=\hat{a}e^{-i\varphi} + \hat{a}^\dag e^{i\varphi}$.
    Here, $\hat{a}^\dagger$ and $\hat{a}$ denote the bosonic creation and annihilation operator and $D(\alpha)=\exp(\alpha\hat{a}^\dag-\alpha^*\hat{a})$ is the displacement.
    For a measurement of the quadrature, the outcome probability in Eq.~\eqref{eq:Bayes} is obtained from the marginals of the Wigner functions $W_{\rho}$ and $W_E$, i.e.,
    \begin{equation}
        \label{eq:prob-Wigner}
        \begin{split}
            p_{\pb}(x_{\varphi})&=\frac{\braket{x_{\varphi}|\rho|x_{\varphi}}\braket{x_{\varphi}|E|x_{\varphi}}}{\int \braket{x_{\varphi}|\rho|x_{\varphi}}\braket{x_{\varphi}|E|x_{\varphi}} dx_{\varphi}},\\
            &=\frac{\int W_{\rho}(x_{\varphi},p_{\varphi}) dp_{\varphi} \int W_{E}(x_{\varphi},p_{\varphi}) dp_{\varphi}}{\int \left(\int W_{\rho}(x_{\varphi},p_{\varphi}) dp_{\varphi} \right)\left(\int W_{E}(x_{\varphi},p_{\varphi}) dp_{\varphi} \right)dx_{\varphi}},\\
        \end{split}
    \end{equation}
    where $\ket{x_{\varphi}}$ is a quadrature eigenstate and the Wigner function $W(\alpha)=W(x_{\varphi},p_{\varphi})$ is parametrized in terms of conjugate quadratures $(x_{\varphi},p_{\varphi})$, with $\hat{p}_{\varphi}=\hat{x}_{\varphi+\pi/2}$.
    For instance, if $\varphi=0$ these are the dimensionless position $\hat{x}_0=\hat{a}+\hat{a}^\dag$ and momentum $\hat{p}_0=i(\hat{a}^\dag-\hat{a})$, and the marginals of the Wigner function $W(x_0,p_0)=W(\mathrm{Re}(\alpha),\mathrm{Im}(\alpha))$ are obtained by integration over the real or imaginary part of $\alpha$. 
    
    Accounting for entanglement with another mode $B$, both in the initial preparation as well as in the postselection, leads to a reduced operator $\Xi_A=\sum_{j,k}x_{jk}\rho^j\otimes E^k$ as in Eq. \eqref{eq:red-past-state}. 
    Evaluating Eq. \eqref{eq:prob-Xi} for a measurement of the quadrature the outcome probabilities become
    \begin{equation}
        p_{\pb}(x_{\varphi})\propto \sum_{j,k}x_{jk}\int W_{\rho^j}(x_{\varphi},p_{\varphi}) dp_{\varphi} \int W_{E^k}(x_{\varphi},p_{\varphi}) dp_{\varphi},
    \end{equation}
    depending on the Wigner functions $W_{\rho^j}$ and $W_{E^k}$.

    \begin{figure*}[t]
        \centering
        \begin{tikzpicture}
        \node at (-4.5,0) {\includegraphics[width=0.49\textwidth]{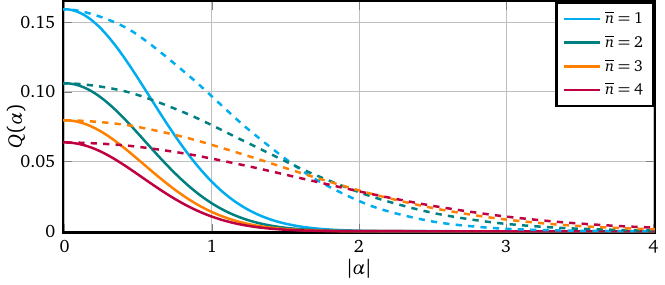}};
        \node at (-2.56,1) {\includegraphics[width=0.115\textwidth]{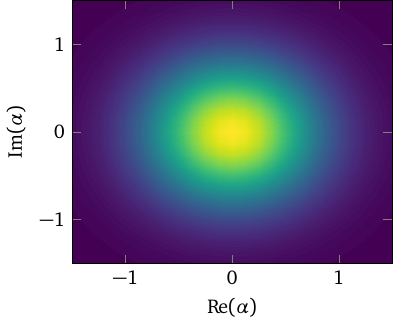}};
        \node at (-4.05,2.2) {(a)};
        \node at (4.5,0.08) {\includegraphics[width=0.49\textwidth]{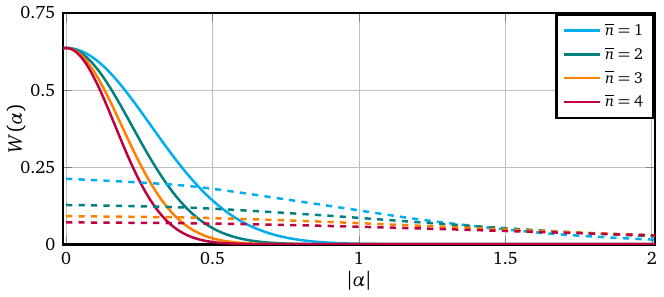}};
        \node at (6.44,1.01) {\includegraphics[width=0.115\textwidth]{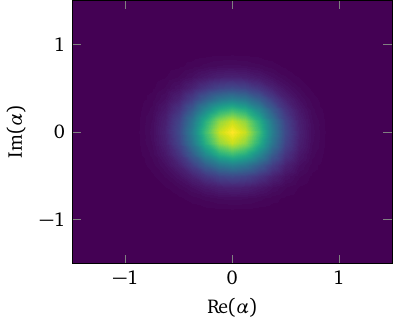}};
        \node at (4.83,2.2) {(b)};
        \end{tikzpicture}
        \caption{\label{fig:phase-space} Husimi function $Q(\alpha)$ (a) and Wigner function $W(\alpha)$ (b) of the statistical operator $\xi$ (solid lines) and the thermal state $\rho_{\mathrm{th}}$ (dashed lines) for different squeezing $\overline{n}=\sinh^2(s)$.
        The densities in the inserts of panels (a) and (b) show $Q_{\xi}(\alpha)$ and $W_{\xi}(\alpha)$, respectively, for $\overline{n}=1$.}
    \end{figure*}

   \subsection{Retrodiction with two-mode squeezed states}  

    As an example, consider a two-mode squeezed state
    \begin{equation}
        \label{eq:squeezed-state}
        \begin{split}
            \ket{\psi_s}&=e^{s(\hat{a}_A^\dag \hat{a}_B^\dag - \hat{a}_A \hat{a}_B)}\ket{00},\\
                &=\frac{1}{\cosh(s)}\sum_{j=0}^\infty \tanh(s)^j\ket{jj},\\
        \end{split}
    \end{equation}
    with squeezing $s$.
    Here, the (photon-)number states $\ket{j}=\tfrac{1}{\sqrt{j!}}(\hat{a}^\dag)^j\ket{0}$ form an orthonormal basis for each mode.
    Suppose $\ket{\psi_s}$ is initially prepared and that postselection is done by projecting onto $\ket{e_s}=\ket{\psi_{-s}}$.
    From Eq. \eqref{eq:red-past-state-general} we obtain the reduced past state as
    \begin{equation}
        \label{eq:red-state-tmsv}
        \Xi_A=\sum_{j,k=0}^\infty \frac{(-1)^{j+k}}{\cosh^2(s)}\tanh(s)^{2(j+k)}\ket{jk}\bra{kj}.
    \end{equation}
    Note that $\Xi_A$ is not positive, and thus not separable, being a sign of the entanglement between $A$ and $B$.
    Using $\Xi_A$ to compute the probabilities in Eq. \eqref{eq:prob-Xi} for a rank-one POVM, $\Omega_m=\ket{\phi_m}\bra{\phi_m}$, yields
    \begin{equation}
        \label{eq:prob-simp}
        p_{\pb}(m)=\frac{|\braket{\phi_m|\xi|\phi_m}|^2}{\sum_{m^\prime}|\braket{\phi_{m^\prime}|\xi|\phi_{m^\prime}}|^2},
    \end{equation}
    where we introduced the single-mode operator
    \begin{equation}
        \label{eq:therm-op}
        \xi=\sum_{j=0}^\infty\frac{(-\overline{n})^j}{(1+\overline{n})^{j+1}}\ket{j}\bra{j},
    \end{equation}
    with $\overline{n}=\sinh^2(s)$.
    The operator $\xi$ has strong resemblance with a thermal state, $\rho_{\mathrm{th}}=\sum_j\tfrac{\overline{n}^j}{(1+\overline{n})^{j+1}}\ket{j}\bra{j}$, with mean photon-number $\overline{n}$, but has an alternating sign in Eq.~\eqref{eq:therm-op}.
    Indeed, if we postselected on the same state $\ket{\psi_s}$, instead of $\ket{\psi_{-s}}$, then the alternating sign in Eq. \eqref{eq:therm-op} vanishes.
    In this case, $\xi$ coincides with the thermal state $\rho_{\mathrm{th}}$ and the outcome probability is $p_{\pb}(m)\propto |\braket{\phi_m|\rho_{\mathrm{th}}|\phi_m}|^2$ corresponding to both the preparation of, and postselection on, a thermal state. 
    
    We emphasize that our formalism allows to calculate outcome probabilities using a single mode operator $\xi$ instead of a pair of two-mode operators $(\psi_s,e_s)$.
    Indeed, mode $B$ and its infinite-dimensional Hilbert space $\mathcal{H}_B$ are completely absent in the evaluation of Eq. \eqref{eq:prob-simp}.

    \subsection{Phase-space representation of $\xi$ and $\rho_{\mathrm{th}}$}

    Given the past state $(\psi_s,e_s)$, the outcome probabilities in Eq. \eqref{eq:gen-Husimi} take the particular simple form $p_{\pb}(\alpha)\propto Q_{\xi}(\alpha)^2$.
    The Husimi function 
    \begin{equation}
        \begin{split}
            Q_\xi(\alpha)&=\frac{e^{-\frac{1+2\overline{n}}{1+\overline{n}}|\alpha|^2}}{\pi(1+\overline{n})},\\
        \end{split}
    \end{equation}
    of $\xi$ is shown in Fig. \ref{fig:phase-space} (a) for different values of $\overline{n}$. 
    For comparison, we also plotted the Husimi function of $\rho_{\mathrm{th}}$,
    \begin{equation}
        Q_\rho(\alpha)=\frac{e^{-\frac{|\alpha|^2}{1+\overline{n}}}}{\pi(1+\overline{n})}.
    \end{equation}
    The Husimi representation of $\rho_{\mathrm{th}}$ and $\xi$ are Gaussian functions, but $Q_{\xi}(\alpha)$ decreases faster than $Q_\rho(\alpha)$.
    Note that in the absence of squeezing, $s=0$, we have $\overline{n}=0$, and both $\rho_{\mathrm{th}}$ and $\xi$ coincide with the vacuum state $\ket{0}$.
    
    Computing the Wigner function of $\xi$ via Eq. \eqref{eq:def-Wigner} gives
    \begin{equation}
        \label{eq:Wigner-xi}
        \begin{split}
            W_{\xi}(\alpha)=\frac{2}{\pi}e^{-2(1+2\overline{n})|\alpha|^2},
        \end{split}
    \end{equation}
    which is shown in Fig.~\ref{fig:phase-space} (b).
    For comparison, we also provide the Wigner function of the thermal state $\rho_{\mathrm{th}}$,
    \begin{equation}
        W_{\rho}(\alpha)=\frac{2}{\pi}\frac{e^{-\frac{2|\alpha|^2}{1+2\overline{n}}}}{1+2\overline{n}}.
    \end{equation}
    Both $\rho_{\mathrm{th}}$ and $\xi$ have Gaussian Wigner functions, but $W_{\xi}(\alpha)$ decreases faster than $W_\rho(\alpha)$.
    In particular, $W_{\xi}(\alpha)$ has variance below the vacuum state and becomes point-like in the limit of infinite squeezing, $\overline{n}\to\infty$.
    This gives an interesting characterization of a result found in Ref.~\cite{KM21}.
    There it was shown that for the past state $(\psi_s,e_s)$ the outcome of any quadrature measurement can be inferred with certainty when given sufficient squeezing.
    That is, $\mathrm{Var}_{\pb}(\hat{x}_{\varphi})\to 0$, for all $\varphi$, when $s\to \infty$.
    In particular, this undercuts the Heisenberg uncertainty relation derived from the commutator $[\hat{x}_\varphi,\hat{p}_{\varphi}]=2i$.
    Here, we obtain the same result by noting that the Wigner function of the normalized operator $\hat{\xi}= \xi/\Tr(\xi)$, with $\Tr(\xi)=1/(1+2\overline{n})$, becomes a Dirac $\delta$ distribution, 
    \begin{equation}
        W_{\hat{\xi}}(x_{\varphi},p_{\varphi})\to\delta(x_{\varphi})\delta(p_{\varphi}),
    \end{equation}
    when $s\to\infty$.
    Then, the outcome $x_{\varphi}$ is indeed obtained with certainty. 
    In contrast, the Wigner function of a density matrix $\rho$ cannot be singular as is embodied by the Heisenberg uncertainty principle.
    Thus it cannot give rise to such determined measurements of the quadrature $\hat{x}_\varphi$ for different angles $\varphi$.
    Notably, Ref. \cite{S16} showed that for $\overline{n}=1/2$ the operator $\xi$ constitutes a limit-case on how singular the phase-space distribution of a density matrix can be.
    For $\overline{n}>1/2$, we find a clear violation of this bound, e.g., in terms of a singular distribution $\delta(\alpha)$.

\section{Dimension witnesses for past states}
    \label{sec:dim-wit}

We have so far shown how entanglement properties of the state and the measurement effect of a composite system $AB$ are revealed by the matrix properties (the eigenvalues) of the statistical operator \eqref{eq:red-past-state-general} that acts only on the system $A$. 
In this section, we will proceed and investigate to what extent the postselected (and retrodicted) outcomes on the subsystem can reveal the nature of the entanglement, and in particular, the Hilbert space dimension of the unobserved part $B$ of the entangled state.  

We assume that we know the dimension of the (accessible) subsystem $A$, and the task is to give a lower bound for the dimension of subsystem $B$. 
To this end, we adapt the notion of dimension witnesses \cite{BPA08,JP11,PM20} to past quantum states. 
Let $\mathcal{D}(\mathcal{H}_A\otimes\mathcal{H}_B)$ denote the set of positive operators on the composite system $AB$.
    A function $f$ is said to be a (proper) dimension witness iff
    \begin{align*}
        \text{(i)}\quad\forall \rho, E\in\mathcal{D}\big(\mathbb{C}^{d}\otimes \mathbb{C}^{k}\big):&\quad f(\bm{P}_{\rho,E})\leq c_k,\\
        \text{(ii)}\quad\exists \rho, E\in\mathcal{D}\big(\mathbb{C}^{d}\otimes \mathbb{C}^{D}\big):&\quad f(\bm{P}_{\rho,E})> c_k,
    \end{align*}
    for some $D>k$.
    Here, $\big(\bm{P}_{\rho,E}\big)_{a,m}=p_{\pb}^{(a)}(m)$ is a matrix, whose entries are the probabilities in Eq. \eqref{eq:Bayes-local}, and the letter $a$ enumerates different possible choices of measurement, i.e., POVM operators $\Omega^{(a)}_{m}$.
    Since $P_{\rho,E}$ depends only on measurements performed on system $A$, without any access to $B$, it is completely determined by the reduced statistical operator $\Xi_{A}$ in Eq. \eqref{eq:red-past-state-general}.
    Thus, one may equivalently write $f(\bm{P}_{\rho,E})=f(\Xi_A)$.
    
    If there does not exist a past state $(\rho,E)$ for which (ii) holds, then we refer to $f$ as a trivial witness.
    The constants $c_k$ give a lower bound, whose violation, i.e., $f(\bm{P}_{\rho,E})>c_k$, signals that the subsystem $B$ must at least be of dimension $k+1$. 
    The definition in (i) implies that
    \begin{equation}
        \label{eq:order}
        c_1\leq c_2 \leq \dots \leq c_d.
    \end{equation}
    Moreover, we have $c_0=c_1$ but it also holds that $c_d=c_{d+1}$, which is shown in Appendix \ref{app:Schmidt} using a Schmidt decomposition of $\rho$ and $E$.
    Then, one only tests for dimensions $D\leq d$ as no larger space $\mathcal{H}_B$ can be detected.

    Whether the witness $f$ is trivial depends on the choice of measurements $\Omega^{(a)}_{m}$  that underlie the probabilities $\bm{P}_{\rho,E}$.
    For example, for a single projective measurement, $\Omega_m=\Pi_m$, we have that $(\bm{P})_m=p_{\pb}(m)$ corresponds to a single probability vector.
    Outcome probabilities of this form can be generated by the reduced density matrix $\rho_A=\sum_m p_{\pb}(m)\Pi_m$.
    In this case, neither retrodiction nor entanglement leave any unique signature to be detected; the witness $f$ is trivial.
    
    In general, the constants $c_k$ are solutions to the optimization problem
    \begin{equation}
        \label{eq:opt-past-mixed}
        c_k=\max_{\rho, E\in\mathcal{D}(\mathbb{C}^{d}\otimes \mathbb{C}^{k})}f(\bm{P}_{\rho,E}).
    \end{equation}
    This amounts to an optimization with $2(d^2k^2-1)$ real-valued parameters necessary to specify the pair $(\rho,E)$.
    At first sight this task appears to be computationally harder than the optimization problems one usually encounters in witness-based approaches to detect dimension and entanglement \cite{LK00,GVS18,PAH24}, because these only involve the density matrix $\rho$.
    However, in the remainder of this section, we show that Eq. \eqref{eq:opt-past-mixed} can be simplified significantly.

    \subsection{Convex dimension witnesses}
    \label{ssec:convex}
    
    The conditional probabilities in Eq. \eqref{eq:Bayes} are nonlinear in both $\rho$ and $E$.
    Despite this, they still obey some generalized form of (quasi-)convexity.
    That is, for convex sums $\rho=\sum_xp_x\ket{\psi_x}\bra{\psi_x}$ and $E=\sum_y q_y\ket{e_y}\bra{e_y}$ there exist new probabilities $\{p_x^\prime\}_x$ and $\{q_{xy}^\prime\}_y$, such that 
    \begin{equation}
        \label{eq:convex-Bayes}
        p_{\pb}(m)=\sum_{x,y}p_x^\prime q_{xy}^\prime p_{\pb}(m|x,y),
    \end{equation}
    where
    \begin{equation}
        p_{\pb}(m|x,y)=\frac{|\bra{e_y}\Omega_m\ket{\psi_x}|^2}{\sum_{m^\prime} |\bra{e_y}\Omega_{m^\prime}\ket{\psi_x}|^2}.
    \end{equation}
    The proof is given in Appendix \ref{app:convex}.
    Now, there are indices $x^\prime$ and $y^\prime$ such that $p_{\pb}(m)\leq p_{\pb}(m|x^\prime,y^\prime)$.
    This result is useful if we assume the witness $f(\bm{P})$ to be a convex function, i.e.,
    \begin{equation}
        f\Big(\sum_{z}r_z\bm{P}_{z}\Big)\leq \sum_{z}r_z f(\bm{P}_{z}),
    \end{equation}
    for matrices $\bm{P}_z$ and probabilities $r_z\geq 0$.
    Then, to determine $c_k$, it suffices to optimize in Eq. \eqref{eq:opt-past-mixed} over pure states and effects only, i.e.,
    \begin{equation}
        \label{eq:opt-past-pure}
        c_k=\max_{\ket{\psi},\ket{e}\in \mathbb{C}^{d}\otimes \mathbb{C}^{k}}f(\bm{P}_{\psi,e}).
    \end{equation}
    For example, a witness given by a matrix norm, $f(\bm{P})=\Vert\bm{P}\Vert$, is convex. 

    Notably, to violate a witness criteria $c_k$, both $\rho$ and $E$ have to be entangled.
    To see this, let, without loss of generality, 
    \begin{equation}
        \label{eq:sep-op}
        \sigma=\sum_{x}p_x \ket{\psi^x}\bra{\psi^x}\otimes \ket{\phi^x}\bra{\phi^x},
    \end{equation}
    be a separable density matrix and $E$ be an arbitrary effect. 
    Suppose the past state $(\sigma,E)$ could violate one of the criteria $c_k$, i.e., $f(\bm{P}_{\sigma,E})>c_k$, for some $k$.
    According to the ordering in Eq. \eqref{eq:order} it must then also hold that $f(\bm{P}_{\sigma,E})>c_0$.
    Using the (quasi-)convexity from Eq. \eqref{eq:convex-Bayes}, there exists a product state $\ket{\psi^x}\otimes\ket{\phi^x}$ in the ensemble \eqref{eq:sep-op} such that $f\big(\bm{P}_{\psi^x\otimes \phi^x,E}\big)>c_0$.
    On the other hand, for $(\psi^x\otimes\phi^x,E)$ the probabilities in Eq. \eqref{eq:Bayes-local} become
    \begin{equation}
        \begin{split}
            p_{\pb}(m)&=\frac{\bra{\psi^x}\Omega_m^\dag E^{x}\Omega_m\ket{\psi^x}}{\sum_{m^\prime}\bra{\psi^x}\Omega_{m^\prime}^\dag E^{x}\Omega_{m^\prime}\ket{\psi^x}},\\
        \end{split}
    \end{equation}
    where $E^x=\braket{\phi^x|E|\phi^x}$ is a reduced effect matrix.
    However, this implies that
    \begin{equation}
        \bm{P}_{\psi^x\otimes\phi^x,E}=\bm{P}_{\psi^x,E^x},
    \end{equation}
    is completely determined by the past state $(\psi^x,E^x)$ on $\mathcal{H}_A=\mathbb{C}^d$.
    For $(\psi^x,E^x)$ we have, by definition (i), $f(\bm{P}_{\psi^x,E^x})\leq c_0$, leaving us with a contradiction.
    
    In conclusion, the dimension witness $f$ is only triggered by entanglement in both $\rho$ and $E$.
    In particular, it is agnostic towards classical correlations between $A$ and $B$.
    From this viewpoint, the constant $c_0$ plays the role of an entanglement witness \cite{GT09,SV13} for the reduced past state $\Xi$ in Eq. \eqref{eq:red-past-state}, i.e., $c_0-f(\Xi)\geq 0$ for all situations where $\Xi$ is separable.

    \subsection{Entanglement classification of past quantum states}
    \label{ssec:ent}
    
    The optimization in Eq. \eqref{eq:opt-past-pure} can be simplified even further. 
    Notice that any state with Schmidt-rank at most $k$ can be written as $\ket{\psi}=(\mathbb{1}\otimes M)\ket{\phi^+_k}$, where $\ket{\phi^+_k}=\tfrac{1}{\sqrt{k}}\sum_{j=1}^{k}\ket{jj}$, is an entangled state with Schmidt-rank $k$, and $M$ encodes the desired state amplitudes \cite{HHH09}. Define the effect as $\bra{e_M}=\bra {e}(\mathbb{1}\otimes M)$, and observe that Eq. \eqref{eq:Bayes-local} yields
    \begin{equation}
        \begin{split}
            p_{\pb}(m)&=\frac{|\bra{e}(\Omega_m\otimes \mathbb{1})\ket{\psi}|^2}{\sum_{m^\prime} |\bra{e}(\Omega_{m^\prime}\otimes \mathbb{1})\ket{\psi}|^2},\\
            &=\frac{|\bra{e_M}(\Omega_m\otimes \mathbb{1})\ket{\phi^+_k}|^2}{\sum_{m^\prime} |\bra{e_M}(\Omega_{m^\prime}\otimes \mathbb{1})\ket{\phi_k^+}|^2}.\\
        \end{split}
    \end{equation}
    This implies that in Eq. \eqref{eq:opt-past-pure} it suffices to optimize over pure effects only, i.e.,
    \begin{equation}
        c_k=\max_{\ket{e}\in \mathbb{C}^{d}\otimes \mathbb{C}^{k}}f\big(\bm{P}_{\phi^+_k,e}\big).
    \end{equation}
    Remarkably, it is only necessary to optimize over $2dk-1$ real-valued parameters specifying the effect $\ket{e}$.

    We note that in the literature on entanglement, the (usually invertible) filter operation $M$ connects states that are convertible under stochastic LOCC (local operations and classical communication) \cite{DVC00,I04}.
    This highlights that the past state $(\rho,E)$ is an inherently postselected state \cite{AV91}, due to the effect $E$.
    Thus, typical entanglement measures being invariant under LOCC, such as distance measures or entropic quantities, are insufficient to characterize the entanglement of past states.
    Instead stochastic-LOCC monotones have to be employed for their characterization, such as the Schmidt number \cite{SV11}, or a more general convex ordering \cite{SV15}.
    
    \section{Measurements of mutually unbiased bases}
    \label{sec:examples}

    In this section, we illustrate the theory of dimension witnesses for projective measurements on mutually unbiased bases, including Pauli measurements.

    \subsection{Ideal Pauli measurements}
    \label{ssec:Pauli}

    As our first example, we consider system $A$ to be a qubit ($d=2$) and the measurements are given by the Pauli matrices $\sigma^x$, $\sigma^y$, and $\sigma^z$.
    The witness $f$ can be represented by the past Bloch vector $\bm{r}_\pb=(\braket{\sigma^x}_{\pb},\braket{\sigma^y}_{\pb},\braket{\sigma^z}_{\pb})$.
    To obtain $c_0$, we use the expansion
    \begin{equation}
        \label{eq:Bloch-exp}
        \begin{split}
            \rho&=\mathbb{1} + r_x\sigma^x + r_y\sigma^y +  r_z\sigma^z,\\
            E&=\mathbb{1} + s_x\sigma^x + s_y\sigma^y +  s_z\sigma^z,\\
        \end{split}
    \end{equation}
    where $r_a=\Tr(\rho\sigma^a)$ and $s_b=\Tr(E\sigma^b)$ are the conventional Bloch-vector components of $\rho$ and $E$, respectively.
    In Eq. \eqref{eq:Bloch-exp} normalization can be neglected, because the probabilities in Eq.~\eqref{eq:Bayes} are independent of the overall scaling of both $\rho$ and $E$.
    The components of $\bm{r}_\pb$ are \cite{TFN17}
    \begin{equation}
            \braket{\sigma^a}_{\pb}=\frac{r_a + s_a}{1 + r_a s_a},
    \end{equation}
    for $a=x,y,z$.
    Next, we choose a witness as
    \begin{equation}
        \label{eq:abs-norm}
        f(\bm{r}_\pb)=|\braket{\sigma^x}_\pb| + |\braket{\sigma^y}_\pb| + |\braket{\sigma^z}_\pb|.
    \end{equation}
    An optimization over the real-valued parameters $r_a$ and $s_b$ reveals that $c_0\approx 2.598$, which is obtained for $\braket{\sigma^a}_{\pb}\approx 0.86$.
    For arbitrary past states on the joint system $AB$, the witness is bounded by $f(\bm{r}_\pb)\leq 3$, because $|\braket{\sigma^a}_{\pb}|\leq 1$.

    As shown in \cite{VAA87}, the past state $(\psi,e)$ from Eq. \eqref{eq:VAA-state} reaches this bound.
    In this case it is possible to retrodict with certainty the outcome of any of the three Pauli measurements, as $\braket{\sigma^x}_{\pb}=\braket{\sigma^y}_{\pb}=\braket{\sigma^z}_{\pb}=1$. 
    By the violation of the dimension criteria, $f(\bm{r}_\pb)=3>c_0$, Pauli measurements on $A$ alone thus suffice to verify the entanglement with another system $B$ in both the preparation and later detection. 
    As further shown in \cite{VAA87}, for all four possible outcomes of measurements on the final state, in a definite orthogonal basis of entangled states, retrodiction with certainty of measurements of any of the three spin components is possible.    
 
\begin{figure}[t]
        \centering
        \begin{tikzpicture}
        \node at (0,0) {\includegraphics[width=0.42\textwidth]{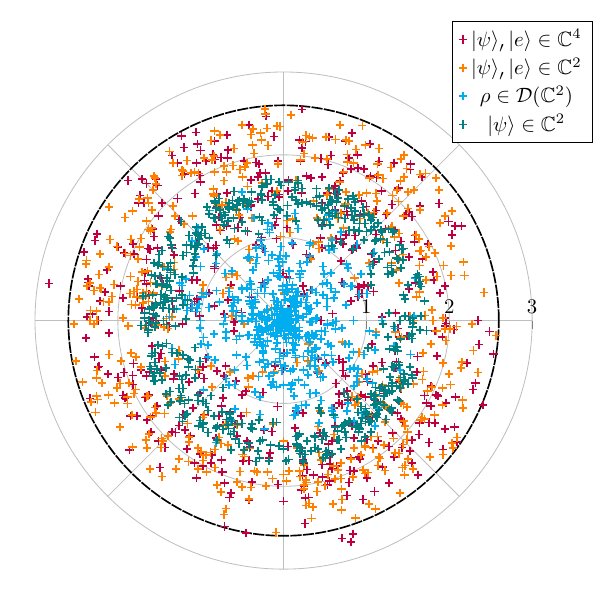}};
        \end{tikzpicture}
        \caption{\label{fig:wit-Pauli} Evaluation of $f(\bm{r}_\pb)$ in Eq. \eqref{eq:abs-norm} for the past Bloch vector $\bm{r}_\pb$ of a Haar-random past state.
        The blue marks correspond to mixed states, green marks belong to pure states, and the orange marks correspond to pure past states on $\mathbb{C}^2$. 
        The red marks belong to pure past states on the larger system $\mathbb{C}^4$.
        The black dashed line corresponds to $c_0\approx 2.598$.}
    \end{figure}

    Figure \ref{fig:wit-Pauli} shows the witness $f$ evaluated on past states sampled uniformly from a Haar-random set of matrices.
    Only entangled past states (red marks) have the potential to lye outside the region marked by $f(\bm{r}_\pb)\leq c_0$.
    Details on the numerical generation of Haar-random matrices are given in Appendix \ref{app:Haar} and in Ref. \cite{M07}.

    In summary, we saw that entangled past states enable the sharp retrodiction of the outcomes of multiple noncommuting observables beyond what can be achieved by separable states.
    Roughly speaking, a useful witness quantifies the degree of certainty that one has about the outcomes of multiple measurements.
    A witness should be composed in a way that does not add redundant information, as this waters down the dimension criteria.

    \subsection{Pauli measurements with uncertain outcomes}
    \label{ssec:noise}

\begin{figure}[t]
        \centering
        \begin{tikzpicture}
        \node at (0,0) {\includegraphics[width=0.5\textwidth]{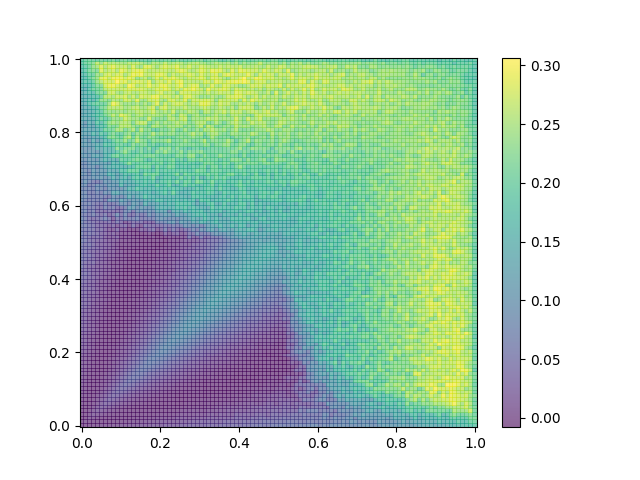}};
        \node at (-0.5,-3.1) {$p$};
        \node at (-4,-0.1) {$q$};
        \node at (2.7,2.8) {$\Delta$};
        \end{tikzpicture}
        \caption{\label{fig:noisy}
        Detection gap $\Delta(p,q)$ as a function of the probabilities $p$ and $q$ for the POVMs in Eq. \eqref{eq:Pauli-noise} and the witness in Eq. \eqref{eq:wit-noise}.}
    \end{figure}
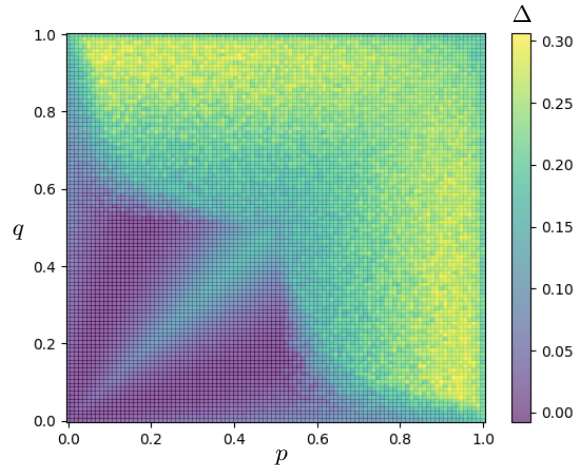

    Next, we consider Pauli measurements which are subject to imperfections, as illustrated by a three outcome measurement on qubits for each of the three spin directions ($a=x,y,z$). 
    The corresponding POVM elements
    \begin{equation}
        \label{eq:Pauli-noise}
    \begin{split}
        \Omega^{(a)}_{0}&=\sqrt{p_a}\,\Pi_{0}^a\\ 
        \Omega^{(a)}_{1}&=\sqrt{q_a}\,\Pi_{1}^a\\
        \Omega^{(a)}_{2}
         &=\sqrt{1-p_a}\Pi_{0}^{a}+ \sqrt{1-q_a}\Pi_{1}^a\\
        \end{split}
    \end{equation}
    are not orthogonal projection operators.
    The outcomes $m=0,1$ verify that the system is in the corresponding eigenstate of the Pauli matrix $\sigma^a$ with success probability $0\leq p_a,q_a\leq 1$. 
    The outcome $m=2$ yields less information about the state (no information if $p_a=q_a$).
    To gauge the impact of the uncertain outcomes on our ability to witness the presence of $B$, we introduce the detection gap $\Delta=c_2-c_0$ ($c_0=c_1$) of the witness
    \begin{equation}
        \label{eq:wit-noise}
        f(\bm{P})=\sum_{a=x,y,z}\max_{m} p_{\pb}^{(a)}(m).
    \end{equation}
    The detection gap $\Delta$ marks the interval in which the presence of $B$ can be detected.  
    We obtain $\Delta$ from the optimization in Eq. \eqref{eq:opt-past-pure} for $c_0$ and $c_2$. 
    Figure \ref{fig:noisy} shows $\Delta(p,q)$ as a function of $p=p_a$ and $q=q_a$.
    The darker shaded regions mark a vanishing detection gap, showing that very noisy measurements, $p,q\approx 0$, reveal little information and do therefore not allow us to verify the presence of $B$.
    Note, however, that $p\approx 1$ and $q\approx 0$ also offer a large detection gap, since our model then makes $\Omega^{(a)}_{2}=\Pi_{1}^a$ the certain projective measurement of an eigenstate of $\sigma^a$. 
    The same argument applies for $p\approx 0$ and $q\approx1$, so that $\Omega^{(a)}_{2}=\Pi_{0}^a$ is again a projection.

    \subsection{Measurements in higher dimensions}

    Consider a qudit ($d=4$) and projective measurements
    \begin{equation}
        \label{eq:MUB-qudit}
         \Pi_{m}^a=U_a \ket{m}\bra{m} U_a^\dag,
    \end{equation}
    with $m=0,1,2,3$ and unitary matrices $U_0=\mathbb{1}$,
    \begin{equation*}
        \begin{split}
            U_1&=\frac{1}{2}\begin{bmatrix}
            1 & 1 & 1 & 1\\
            1 & 1 & -1 & -1\\
            1 & -1 & -1 & 1\\
            1 & -1 & 1 & -1\\
            \end{bmatrix},\quad U_2=\frac{1}{2}\begin{bmatrix}
            1 & 1 & 1 & 1\\
            -1 & -1 & 1 & 1\\
            -i & i & i & -i\\
            -i & i & -i & i\\
            \end{bmatrix},\\
        \end{split}
    \end{equation*}
    and 
    \begin{equation*}
        U_3=\frac{1}{2}\begin{bmatrix}
            1 & 1 & 1 & 1\\
            -i & -i & i & i\\
            -i & i & i & -i\\
            -1 & 1 & -1 & 1\\
            \end{bmatrix},\quad U_4=\frac{1}{2}\begin{bmatrix}
            1 & 1 & 1 & 1\\
            -i & -i & i & i\\
            -1 & 1 & -1 & 1\\
            -i & i & i & -i\\
            \end{bmatrix}.
    \end{equation*}
    These measurements constitute mutually unbiased bases \cite{KR04}, i.e., $|\braket{m|U_{a}^\dag U_b|n}|^2=\tfrac{1}{4}$, for $a\neq b$.
    
\begin{figure}[t]
        \centering
        \begin{tikzpicture}
        \node at (0,0) {\includegraphics[width=0.4\textwidth]{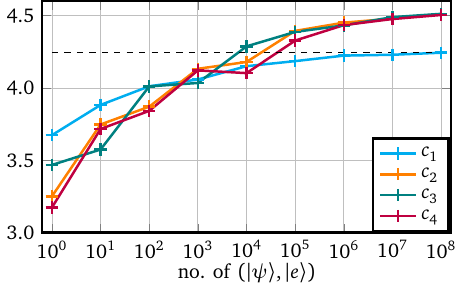}};
        \end{tikzpicture}
        \caption{\label{fig:bounds}
        Numerical estimate of the dimension criteria $c_k$ in Eq. \eqref{eq:opt-past-pure} for the witness given in Eq. \eqref{eq:wit-qudit} and the mutually unbiased basis in Eq. \eqref{eq:MUB-qudit}.
        The estimate is obtained from a Haar-random sampling of (pure) past qudit states $\ket{\psi},\ket{e}\in\mathbb{C}^4\otimes\mathbb{C}^k$, for $k=1,2,3,4$ and for a varying sample size.}
    \end{figure}

    We choose, similar to Eq. \eqref{eq:wit-noise}, the witness as
    \begin{equation}
        \label{eq:wit-qudit}
        f(\bm{P})=\sum_{a=0}^4\max_{m}p_{\pb}^{(a)}(m).
    \end{equation}
    The constants $c_k$ from Eq. \eqref{eq:opt-past-pure} are estimated by evaluating the witness $f(\bm{P}_{\rho,E})$ on a large sample of Haar-random past states $\rho,E\in\mathcal{D}(\mathbb{C}^d\otimes \mathbb{C}^k)$, for $k=1,2,3,4$.
    This brute force approach has the advantage of not needing to parametrize the past state $(\rho,E)$ nor do we need to specify the derivatives of the objective function $f$.
    To gauge the improvement of the optimization by using larger samples of Haar-random matrices, Fig. \ref{fig:bounds} shows the obtained $c_k$ in terms of the sample size.
    We observe that the witness is non-trivial, $c_1< c_2$, so that the presence of (entanglement with) system $B$ can be detected. 
    Notably, our numerical search suggests $c_2\approx c_3\approx c_4$.
    Thus, the POVMs in Eq.~\eqref{eq:MUB-qudit} cannot reliably distinguish if $B$ consists of a qubit or a higher-dimensional state.

\section{Conclusion}
\label{sec:Fin}

In this article we showed that comparing retrodicted outcomes with measurements on a single subsystem can witness entanglement with another system in the initial state and later measurements.
This situation is fundamentally different from what one finds making predictions with Born's rule, and for which entanglement with an ancillary system is fully accounted for by a reduced density matrix, and hence indiscernible from classical randomness in the preparation of the monitored system. 

To account for this phenomenon, criteria for such entanglement were formulated, and we found it useful to introduce a statistical operator acting on the tensor product space of the small system with itself.
Negative eigenvalues of that operator point to the existence and entanglement with environment degrees of freedom, This analysis interestingly singled out bound entanglement and hinted at the possibility that free entanglement might be necessary for the sharp retrodiction of multiple noncommuting observables.
This reduced past state was readily adapted to account for continuous variable entanglement in bosonic systems, reducing the computational effort necessary to obtain conditional probabilities in both homodyne and heterodyne measurements. 
    
Moreover, we devised a theory of dimension witnesses for past quantum states in which a lower bound on the dimension of the environment is derived from outcome probabilities on the accessible system. 
This generalizes the existing theory of dimension witnesses to include Bayesian inference.
Despite the nonlinear form of Bayes rule, we saw that determining these lower bounds amounts to a (quasi-)convex optimization problem.

We illustrated our approach on measurements of mutually unbiased bases, for which we determined the dimension criteria and their violation by higher-dimensional states. Predicting the outcomes of measurements of mutually unbiased bases using postselection is a primary aspect of the mean king's problem \cite{HHH05,RW07,JGA23}. The fact that retrodiction does not follow alone from reduced subsystem properties makes it an interesting topic to discuss in connection with contextuality \cite{S05,T07} and measurement incompatibility \cite{GBC17}.

Our reduced statistical operator provides useful separability criteria and differs from retrodiction in terms of postselected closed-time-like curves \cite{S11,LMG11,SSG24}.
It might be a fruitful endeavor to connect the separability of the statistical operator $\Xi_A$ to different concepts  of temporal entanglement, see, e.g., Refs. \cite{MFC23,MAP25} for recent accounts.

Future research may use aspects of this work to investigate whether a communication channel in a quantum network is noisy due to environmental interactions \cite{MCD93}, the presence of an eavesdropper \cite{AGM06}, or due to interference from a notorious network provider that limits the transmission of quantum resources by quantum censorship \cite{PS24,PM24}.

\acknowledgements
We gratefully acknowledge financial support from Denmarks Grundforskningsfond (DNRF 139, Hy-Q Center for
Hybrid Quantum Networks) and the Alexander von Humboldt Foundation (Feodor Lynen Research Fellowship).

	\appendix

    \section{Schmidt decomposition of past states}
    \label{app:Schmidt}

    Given a witness $f$, the constants $c_k$ give a lower bound on the dimension of a system $\mathcal{H}_B=\mathbb{C}^D$ using only the conditional probabilities in Eq.~\eqref{eq:Bayes-local} given by measurements on system $\mathcal{H}_A=\mathbb{C}^d$.
    Here, we show that these dimension criteria cannot distinguish between $B$ being of dimension $D=d$ and $D>d$.
    
    To see this, note that for $D\geq d$, both $\rho$ and $E$ have Schmidt decompositions
    \begin{equation}
        \ket{\psi}=\sum_{a=0}^{d-1} \ket{\psi_a}\otimes \ket{\hat{\psi}_a},\quad \ket{e}=\sum_{b=0}^{d-1} \ket{e_b}\otimes \ket{\hat{e}_b},
    \end{equation}
    where $\ket{\psi_a},\ket{e_b}\in\mathbb{C}^d$ and $\ket{\hat{\psi}_a},\ket{\hat{e}_b}\in\mathbb{C}^D$.
    Inserting this expansion into Eq. \eqref{eq:Bayes-local} yields 
    \begin{equation}
        \begin{split}
            p_{\pb}(m)
            &=\frac{\big|\sum_{a,b}\braket{\psi_a|\Omega_m|e_b}\braket{\hat{\psi}_a|\hat{e}_b}\big|^2}{\sum_{m^\prime}\big|\sum_{a,b}\braket{\psi_a|\Omega_{m^\prime}|e_b}\braket{\hat{\psi}_a|\hat{e}_b}\big|^2}.\\
        \end{split}
    \end{equation}
    The components of any complex-valued $d\times d$ matrix $M$ can be written as $M_{ab}=\braket{\psi_a^\prime|e_b^\prime}$, with linearly independent vectors $\ket{\psi_a^\prime},\ket{e_b^\prime}\in\mathbb{C}^d$.
    Choosing the $d^2$ complex coefficients $\braket{\hat{\psi}_a|\hat{e}_b}$ to form the matrix $M$ implies $\braket{\hat{\psi}_a|\hat{e}_b}=\braket{\psi_a^\prime|e_b^\prime}$ for all $a,b=0,\dots,d-1$.
    Thus, for any past state $\rho,E\in\mathcal{D}(\mathbb{C}^d\otimes \mathbb{C}^D)$ with $D>d$, there exists a lower-dimensional past state $\rho^\prime,E^\prime\in\mathcal{D}(\mathbb{C}^d\otimes \mathbb{C}^d)$, giving rise to the same probabilities on $A$ via Eq. \eqref{eq:Bayes-local}.
    Hence, $f(\bm{P}_{\rho,E})=f(\bm{P}_{\rho^\prime,E^\prime})$.
    It follows that $c_d=c_D$, for $D\geq d$, thus proving the assertion.

    \section{Generalized convexity of Bayes rule}
    \label{app:convex}

    Despite the non-linearity of the probabilities $p_{\pb}(m)$ in Eq. \eqref{eq:Bayes} in both $\rho$ and $E$, a generalized form of convexity can still be derived. 
    Let $\{\Omega_m\}_m$ be a POVM and express $\rho$ through an ensemble of pure states, $\rho=\sum_x p_x \ket{\psi^x}\bra{\psi^x}$.
    For the probabilities $p_{\pb}(m)$ in Eq. \eqref{eq:Bayes} we then get
    \begin{equation}
        \begin{split}
            p_{\pb}(m)&= \frac{\sum_{x} p_x \alpha(m|x)}{\sum_{m^\prime,x^\prime} p_{x^\prime} \alpha(m^\prime|x^\prime)},\\
        \end{split}
    \end{equation}
    where we defined $\alpha(m|x)=\braket{\psi^x|\Omega^\dag_m E\Omega_m|\psi^x}$.
    Introducing new quantities 
    \begin{equation}
        \begin{split}
            \tilde{p}_x&=\tilde{p} p_x\sum_m \alpha(m|x),\quad
            \tilde{p}= \sum_x\frac{\tilde{p}_x}{\sum_m \alpha(m|x)},
        \end{split}
    \end{equation}
    we rewrite the conditional probabilities as
    \begin{equation}
        \label{eq:gen-convex}
        \begin{split}
            p_{\pb}(m)&=\sum_{x}\frac{\tilde{p}_x}{\sum_{y}\tilde{p}_y}\frac{\alpha(m|x)}{\sum_{m^\prime} \alpha(m^\prime|x)},\\
            &=\sum_{x}p^\prime_x p_{\pb}(m|x),\\
        \end{split}
    \end{equation}
    where $p^\prime_x=\tfrac{\tilde{p}_x}{\sum_{y}\tilde{p}_y}$ are new probabilities and 
    \begin{equation}
        p_{\pb}(m|x)=\frac{\braket{\psi^x|\Omega^\dag_m E\Omega_m|\psi^x}}{\sum_{m^\prime} \braket{\psi^x|\Omega^\dag_{m^\prime} E\Omega_{m^\prime}|\psi^x}}
    \end{equation}
    is a conditional probability with past state $(\psi^x,E)$.

    Without retrodiction, $E=\mathbb{1}^{\otimes 2}$, we have $p_x^\prime=p_x$ and Eq. \eqref{eq:gen-convex} reduces to
    \begin{equation}
        \label{eq:Born-convex}
        p(m)=\sum_xp_xp(m|x),
    \end{equation}
    where $p(m)$ is given by Born's rule, $p(m)=\Tr\big(\Omega^\dag_m\Omega_m \rho\big)$, and $p(m|x)=\braket{\psi^x|\Omega^\dag_m\Omega_m|\psi^x}$.
    In contrast to Eq. \eqref{eq:gen-convex}, Eq. \eqref{eq:Born-convex} amounts to a proper convexity relation.
    
    A similar argument is made for $E=\sum_y q_y \ket{e^y}\bra{e^y}$.
    Define $\alpha(m|x,y)=|\braket{e^y|\Omega_m|\psi^x}|^2$ to obtain
    \begin{equation}
        p_{\pb}(m|x)=\frac{\sum_y q_y\alpha(m|x,y)}{\sum_{m^\prime,y^\prime} q_{y^\prime} \alpha(m^\prime|x,y^\prime)}.
    \end{equation}
    Further introduce 
    \begin{equation*}
        \tilde{q}_{xy}=\tilde{q}_x q_y\sum_{m} \alpha(m|x,y),\quad
            \tilde{q}_x= \sum_y\frac{\tilde{q}_{xy}}{\sum_m \alpha(m|x,y)},
    \end{equation*}
    to obtain 
    \begin{equation}
        \begin{split}
            p_{\pb}(m|x)&=\sum_{y}\frac{\tilde{q}_{xy}}{\sum_{z}\tilde{q}_{xz}}\frac{\alpha(m|x,y)}{\sum_{m^\prime} \alpha(m^\prime|x,y)},\\
            &=\sum_{y}q_{xy}^\prime p_{\pb}(m|x,y),\\
        \end{split}
    \end{equation}
    where $q^\prime_{xy}=\tfrac{\tilde{q}_{xy}}{\sum_{z}\tilde{q}_{xz}}$ are probabilities and 
    \begin{equation}
        p_{\pb}(m|x,y)=\frac{|\braket{e^y|\Omega_m|\psi^x}|^2}{\sum_{m^\prime} |\braket{e^y|\Omega_{m^\prime}|\psi^x}|^2}
    \end{equation}
    is a probability depending on the past state $(\psi_x,e_y)$.

    In summary, we have shown that for any past state $(\rho,E)$, there exists probability distributions $\{p^\prime_x\}_x$ and $\{q^\prime_{xy}\}_y$, such that the conditional probabilities in Eq. \eqref{eq:Bayes} can be written as a convex combination, viz.
    \begin{equation}
        p_{\pb}(m)=\sum_{x,y}p_x^\prime q^\prime_{xy}p_{\pb}(m|x,y).\\
    \end{equation}
    Thus, there ares indices $x^\prime$ and $y^\prime$ for which
    \begin{equation}
        \forall m: \quad p_{\pb}(m)\leq p_{\pb}(m|x^\prime,y^\prime).
    \end{equation}
    It is this quasi-convexity that allows us to restrict the optimization in Eq. \eqref{eq:opt-past-mixed} to pure past states.

    \section{Haar-random matrices}
    \label{app:Haar}

    Given a witness $f$, we determine $c_k$ by solving the optimization problem in Eq.~\eqref{eq:opt-past-pure}.
    A brute force way to achieve this, evaluates $f$ on a large sample of uniformly distributed Haar-random matrices $\rho,E\in\mathcal{D}(\mathbb{C}^d\otimes\mathbb{C}^k)$.
    As pointed out in Sec. \ref{sec:dim-wit}, the optimization only needs to be performed over pure states, i.e., rank-one matrices.
    
    First, we construct Haar-random unitaries using the algorithm from Ref. \cite{M07}. 
    Let $M=QR$ be a $QR$-decomposition of a complex-valued $dk\times dk$ matrix $M$, where $Q$ is unitary and $R$ is an upper triangular matrix.
    Define a diagonal matrix $D = \mathrm{diag}(R_{aa}/|R_{aa}|)$.
    Then, $U=Q D$ is a Haar-random unitary matrix \cite{M07}.
    Applying a Haar-random unitary $U$ to a randomly selected basis vector $\ket{a}$, for $a=0,\dots,dk-1$, results in a Haar-random pure state $U\ket{a}$.
    Let $\mathcal{S}_k\subset\mathcal{D}(\mathbb{C}^d\otimes\mathbb{C}^k)$ be a large sample of Haar-random past states $\ket{\psi}$ and $\ket{e}$ obtained in this way.
    Evaluating $f$ on $\mathcal{S}_k$ provides an approximate solution to the optimization problem,
    \begin{equation}
        c_k \approx \max_{\ket{\psi},\ket{e}\in \mathcal{S}_k} f\big(\bm{P}_{\psi,e}\big),
    \end{equation}
    where $\bm{P}_{\psi,e}$ is given by Eq.~\eqref{eq:Bayes-local}.
    Clearly, the thus obtained approximation gives a lower bound to the actual constant $c_k$, and coincides with $c_k$ in the limit of infinitely long sampling.
    Due to its brute force nature, this approach does not rely on computing the derivative or specifying the Jacobian of the witness $f(\bm{P}_{\rho,E})$ as a function of a parametrization of $(\rho,E)$.

\end{document}